\let\Vec\vec
\documentclass[graybox]{svmult}

\usepackage{mathptmx}       % selects Times Roman as basic font
\usepackage{helvet}         % selects Helvetica as sans-serif font
\usepackage{courier}        % selects Courier as typewriter font
\usepackage{makeidx}         % allows index generation
\usepackage{graphicx}        % standard LaTeX graphics tool
\usepackage{multicol}        % used for the two-column index
\usepackage[bottom]{footmisc}% places footnotes at page bottom

\makeindex             % used for the subject index
\newcommand{\Lie}{\pounds}
\newcommand{\Div}{\mathrm{div}}
\newcommand{\VF}[1]{\kern1.5pt\Vec{\kern-1.5pt\mbox{\boldmath$#1$}}}
\newcommand{\Hat}[1]{\kern1pt\hat{\kern-1pt\mbox{\boldmath$#1$}}}
\newcommand{\rr}{\VF r}

\begin{document}

\title*{Piecewise Conserved Quantities}
%\titlerunning{Short Title}
\author{Tevian Dray}
%\authorrunning{Short Author}
\institute{Tevian Dray
\at Department of Mathematics, Oregon State University, Corvallis, OR 97331, USA,
\\\email{tevian@math.oregonstate.edu}
%\and Second Author \at Institute \email{name@email.address}
}

\maketitle

\abstract{We review the treatment of conservation laws in spacetimes that are
glued together in various ways, thus adding a boundary term to the usual
conservation laws.  Several examples of such spacetimes will be described,
including the joining of Schwarzschild spacetimes of different masses, and the
possibility of joining regions of different signatures.  The opportunity will
also be taken to explore some of the less obvious properties of Lorentzian
vector calculus.}

\section{Introduction}
\label{TDintro}

In 1987, my wife (Corinne Manogue) and I found ourselves in India for 6
months, where we were both Indo-American Fellows.  The highlight of our visit
was the 3 months we spent at TIFR, working with Paddy and others in the
Theoretical Astrophysics Group.  During this visit, Paddy and I wrote a paper
on piecewise Killing vectors~\cite{Piece}, bringing additional mathematical
clarity to the intriguing results I had previously obtained with
't~Hooft~\cite{Shell} on shells of matter in Schwarzschild spacetimes.  Little
did I know at the time that this theme would recur in my research in a variety
of contexts (and more than a dozen papers) over the next 20 years.

As I congratulate Paddy on the occasion of this Festschrift, it is with great
pleasure that I look back on this period near the beginning of our careers.
It seems only fitting that I use this opportunity to summarize my own journey
through several quite different applications of piecewise conserved
quantities.

In Sections~\ref{TDpiece} and~\ref{TDconserve}, I lay out the framework for
analyzing piecewise structures, then in Section~\ref{TDpatch} provide the main
mathematical result, the Patchwork Divergence Theorem, generalizing my work
with Paddy~\cite{Piece}.  The two basic applications are considered next,
namely shells of matter in Section~\ref{TDshells}, and signature change in
Section~\ref{TDsig}.  In Section~\ref{TDcalc}, I point out how the underlying
framework of the Patchwork Divergence Theorem also provides insight into
vector calculus in Lorentzian geometry, providing access to these ideas for
non-experts, including undergraduates.  Finally, a very brief summary is given
in Section~\ref{TDsum}.

\section{Piecewise Smooth Tensors}
\label{TDpiece}

\index{distributional curvature}%
\index{piecewise smooth tensors}%
The long history of the distributional curvature due to piecewise smooth
metric tensors is summarized in~\cite{Junction}.  As discussed there, the
basic setup is two smooth manifolds $M^\pm$ joined along a (possibly null)
hypersurface $\Sigma$, with smooth metric tensors $g^\pm_{ab}$ on either side.
Introducing a step function $\Theta$ which is $0$ on $M^-$ and $1$ on $M^+$,
the metric on $M=M^-\cup M^+$ is given by
\begin{equation}
g_{ab} = (1-\Theta)\, g^-_{ab} + \Theta\, g^+_{ab} .
\end{equation}
One traditionally assumes that the metric is continuous at $\Sigma$,
\begin{equation}
[g_{ab}] := g^+_{ab}\big|_\Sigma - g^-_{ab}\big|_\Sigma = 0 ,
\label{dgdef}
\end{equation}
in which case the connection $\Gamma^c{}_{ab}$ is at worst discontinuous and,
as discussed briefly in~\cite{Shell}, it is straightforward to compute
distributional curvature tensors.  For example, the distributional Ricci
tensor is given by
\begin{eqnarray}
R_{ab}
  &=& (1-\Theta)\, R^-_{ab} + \Theta\, R^+_{ab}
	+ \delta_c[\Gamma^c{}_{ab}] - \delta_b[\Gamma^c{}_{ac}]
\nonumber\\
  &=& (1-\Theta)\, R^-_{ab} + \Theta\, R^+_{ab} + \delta \rho_{ab} ,
\end{eqnarray}
where
\begin{equation}
\delta_c = \delta n_c = \nabla_c\Theta ,
\label{delta}
\end{equation}
so that $n_c$ is normal to $\Sigma$, and the distribution $\delta$ can be
thought of as a Dirac delta function.  However, as shown in~\cite{Junction},
it is enough for the \textit{pullbacks} of $g^\pm_{ab}$ to agree on $\Sigma$
in order to have a well-defined tangent space on all of $M$, which we will
exploit in Section~\ref{TDsig}.

More generally, we can consider other piecewise smooth tensors on $M$, such as
a vector field of the form
\begin{equation}
\xi^a = (1-\Theta)\, \xi_-^a + \Theta\, \xi_+^a
\label{dxidef}
\end{equation}
and its discontinuity $[\xi^a]$, defined in analogy with~(\ref{dgdef}).

\section{Piecewise Conserved Quantities}
\label{TDconserve}

\index{Killing vector}%
As is well-known, a \textit{Killing vector} $\xi^a$ can be contracted with
the stress-energy tensor $T_{ab}$ to yield a conserved quantity
\begin{equation}
X^a = T^{ab} \xi_b
\end{equation}
satisfying
\begin{equation}
\oint X^a N_a dS = 0
\label{conserve}
\end{equation}
where $S$ is any closed, piecewise smooth hypersurface (assumed for the
moment to be nowhere null) with unit normal vector $N^a$.  The vanishing of
this integral is a consequence of the Divergence Theorem, since
\index{Divergence Theorem}%
\begin{equation}
\nabla_a X^a
  = \nabla_a (T^{ab}\xi_b)
  = (\nabla_a T^{ab})\xi_b + T^{ab} \nabla_a \xi_b = 0 ;
\label{divX}
\end{equation}
the first term vanishes by energy conservation, and the second by Killing's
equation.

\index{piecewise Killing vector}%
As defined in~\cite{Piece}, a \textit{piecewise Killing vector} is a piecewise
smooth vector field $\xi^a$ on $M$ of the form~(\ref{dxidef}), such that
$\xi^a_\pm$ are Killing vectors on $M^\pm$.  However, piecewise Killing
vectors are \textit{not} in general Killing vectors on $M$, since
\begin{equation}
\nabla_{(a}\xi_{b)} = \left[\xi_{(a}\right]\delta_{b)}
\end{equation}
which is nonzero if $\xi^a$ is discontinuous.  Referring to~(\ref{divX}), we
see that
\begin{equation}
\nabla_a (T^{ab}\xi_b) = \left[T^{ab} \xi_a\right]\delta_b .
\label{Txi}
\end{equation}
\index{Darmois junction conditions}%
If $\Sigma$ is a spacelike hypersurface, and if the Darmois junction
conditions (continuity of both the intrinsic metric and the extrinsic
curvature) are satisfied there, then the stress-energy tensor is continuous at
$\Sigma$ ($[T^{ab}]=0$), and~(\ref{Txi}) reduces to
\begin{equation}
\nabla_a (T^{ab}\xi_b) = T^{ab} \left[\xi_a\right]\delta_b .
\end{equation}
A natural condition on $\xi^a$ is for its tangential components to agree on
$\Sigma$, in which case
\begin{equation}
\left[\xi_a\right] = \Xi n_a
\label{xicond}
\end{equation}
for some function $\Xi$ defined on $\Sigma$.  Given~(\ref{xicond}), and
using~(\ref{delta}), we will obtain a conserved quantity so long as
\begin{equation}
T^{ab} n_a n_b = 0
\label{Tcond}
\end{equation}
on $\Sigma$.  If $\Sigma$ is spacelike,~(\ref{Tcond}) asserts that the energy
density at $\Sigma$ seen by an observer orthogonal to $\Sigma$ must vanish.

\section{The Patchwork Divergence Theorem}
\label{TDpatch}

\index{divergence}%
\index{volume element}%
\index{Lie derivative}%
\index{Stokes' Theorem}%
In the language of differential forms, the \textit{divergence} of a vector
field $X$ is defined in terms of the volume element $\omega$ as
\begin{equation}
\Div(X)\,\omega := \Lie_X\omega
\end{equation}
where $\Lie$ denotes the Lie derivative.  Using Stokes' Theorem in the form
\begin{equation}
\oint_{\partial W} \alpha = \int_W d\alpha
\label{Stokes}
\end{equation}
and the identity
\begin{equation}
\Lie_X \alpha = d(i_X\alpha)+i_X(d\alpha) ,
\end{equation}
where $i$ denotes the interior product, one obtains the Divergence Theorem in
the form
\index{interior product}%
\index{Divergence Theorem}%
\begin{equation}
\int_W \Div(X)\,\omega = \oint_S i_X \omega ,
\end{equation}
where $S=\partial W$.  Any 1-form $m$ orthogonal to $S$ determines a unique
volume element $\sigma$ on $S$ through the requirement that
\begin{equation}
m\wedge \sigma = \omega ;
\end{equation}
$\sigma$ is compatible with the induced orientation on $W$ precisely when $m$
is outward pointing.  Using the properties of the interior product, the
Divergence Theorem becomes
\begin{equation}
\int_W \Div(X)\,\omega = \oint_S m(X)\,\sigma .
\label{DivThm}
\end{equation}
So long as $S$ is not null, the right-hand side of~(\ref{DivThm}) is the same
as the integral in~(\ref{conserve}) after obvious identifications.

\index{Patchwork Divergence Theorem}%
For piecewise smooth tensors, we can apply~(\ref{DivThm}) separately on
$M^\pm$ and then add the results.  Given a region $W=W^+\cup W^-$ overlapping
$\Sigma$, we let $S=\partial W$ and $S_0=W\cap\Sigma$.  We can extend $m$ to
outward-pointing 1-forms $m_\pm$ orthogonal to $S^\pm=\partial W^\pm$; on
$\Sigma$ we have $m_-=-m_+=:m_0$.
The \textit{Patchwork Divergence Theorem}~\cite{Patch} then takes the form
\begin{equation}
\int_W \Div(X)\,\omega = \oint_S m(X)\,\sigma - \int_{S^0} m_0([X]) \,\sigma^0 .
\label{PDT}
\end{equation}
Our convention is that $m_0$ is the \textit{1-form} pointing from $M_-$ to
$M_+$; which way the physically equivalent vector field points depends on
whether $\Sigma$ is spacelike, timelike, or null.

\section{Shells of Matter}
\label{TDshells}

\index{shells of matter}%
\index{two-body problem}%
\index{colliding shells}%
A simple model for matter falling into a black hole consists of spherical
shells of massless matter.  Remarkably, as shown originally by
't~Hooft~\cite{Shell}, this situation can be described by an exact solution of
the Einstein field equation, at least in the context of piecewise smooth
tensors.  The special case of a single massless particle sitting at the
horizon of a Schwarzschild black hole~\cite{Shock} remains the only explicitly
known exact solution in general relativity that describes a test particle
moving in the field of another object, and is in this sense the only known
solution to the relativistic two-body problem.  These models have been
generalized to charged black holes~\cite{Charged}, to colliding
shells~\cite{Plane,Shell,Bounce}, and, more recently, to shells of negative
energy~\cite{Negative}.

\index{massless dust}%
In~\cite{Piece}, we considered two Schwarzschild spacetimes with different
masses joined along a null cylinder $\Sigma=\{u=\alpha\}$ representing a
spherical shell of massless dust. The corresponding metric is
\begin{equation}
ds^2 = \cases{
	-\frac{32m^3}{r}e^{-r/2m}du\,dv + r^2\,d\Omega^2 &$(u\le\alpha)$\cr
	-\frac{32m^3}{r}e^{-r/2M}dU\,dV + r^2\,d\Omega^2 &$(u\ge\alpha)$
  }
\end{equation}
where $U$ and $V$ are functions (only) of $u$ and $v$, respectively, and
\begin{eqnarray}
uv &= - \left( \frac{r}{2m}-1 \right) e^{r/2m} &\qquad (u\le\alpha) ,\\
UV &= - \left( \frac{r}{2M}-1 \right) e^{r/2M} &\qquad (u\ge\alpha) .
\end{eqnarray}
Continuity of the metric requires that on $\Sigma$ we have
\begin{equation}
\frac{\alpha}{m} = \frac{U(\alpha)}{MU'(\alpha)} =: \gamma ,
\end{equation}
which implies that
\begin{equation}
\frac{u\partial_u}{m} = \frac{U\partial_U}{M}
\end{equation}
on $\Sigma$.  The only nonzero component of the stress-energy tensor is
\begin{equation}
T_{uu} = \frac{\delta}{\gamma\pi r^2} \>(M-m) ,
\end{equation}
and we have the piecewise Killing vector
\begin{equation}
\xi
  = (1-\Theta)\, \frac{v\partial_v-u\partial_u}{4m}
	+\Theta\, \frac{V\partial_V-U\partial_U}{4M} .
\end{equation}
Thus, $[\xi]$ is proportional to $\partial_V$, satisfying
condition~(\ref{xicond}), while~(\ref{Tcond}) is satisfied by virtue of the
double-null form of the stress-energy tensor.

We therefore obtain an integral conservation law of the form~(\ref{conserve}).
Since the support of $T_{ab}$ is on $\Sigma$, we obtain a conserved quantity
$Q$ by evaluating this integral over any hypersurface $S$ intersecting
$\Sigma$ only once.  We choose
\begin{equation}
S = \cases{
	\{t=\hbox{const}\} &\quad $(u\le\alpha)$ \cr
	\{T=\hbox{const}\} &\quad $(u\ge\alpha)$
  }
\end{equation}
where $t$ and $T$ denote Schwarzschild time in the regions $u\le\alpha$ and
$u\ge\alpha$, respectively, and where the constants are chosen so that
$\Sigma$ is continuous.  Putting this all together, we have
\begin{equation}
-Q = \int_S \left( (1-\Theta)\,T^t{}_t + \Theta\,T^T{}_T \right) dS
\label{Qint}
\end{equation}
which appears to involve the distributional product $\delta\Theta$.  However,
since
\begin{equation}
\frac{\partial u}{\partial r} = \frac{u}{4m} \frac{1}{1-r/2m} ,
\end{equation}
it turns out that
\begin{equation}
T^t{}_t = T^T{}_T = -\frac{\delta(r-r_0)}{4\pi r^2} (M-m)
\end{equation}
where $r_0$ is the radius of the shell where $S$ intersects $\Sigma$.  Thus,
there is no actual step function present in the integrand
in~(\ref{Qint}). Finally, evaluating the integral leads to
\begin{equation}
Q=M-m
\end{equation}
which shows that the energy of the shell is precisely the difference of the
two Schwarzschild masses, as expected.

Thus, the results of~\cite{Piece} can be regarded as an application of the
Patchwork Divergence Theorem in this setting.

\section{Signature Change}
\label{TDsig}

\index{signature change}%
\index{tensor distributions}%
\index{distributional field equations}%
``Spacetimes'' combining both Lorentzian and Euclidean regions were proposed
independently by George Ellis's group in the context of early universe
cosmology~\cite{Ellis1,Ellis2} and by our group in the context of quantum
field theory in curved space~\cite{Change,Scalar,Boundary}.  Subsequent work
by both groups addressed tensor distributions~\cite{Tensor,Tensor2} and the
distributional field equations~\cite{Einstein,Gravity}, in the process
realizing that conservation laws would take a different form at a change of
signature~\cite{Failure}, ultimately leading to the Patchwork Divergence
Theorem~\cite{Patch}.

The key point is that, even though the metric is clearly discontinuous
at a change of signature,
\footnote{We assume the metrics $g^\pm_{ab}$ are non-degenerate on $\Sigma$,
the only other possibility.}
the same need not be true for the volume element.  The easiest way to see this
surprising fact is to construct orthonormal frames on both $M^\pm$, and
compare them along $\Sigma$.  We assume that $\Sigma$ is spacelike as seen
from both sides, in which case an orthonormal frame on $\Sigma$ can be
(separately) extended to orthonormal frames on $M^\pm$ by adding the
appropriate normal vector, which is spacelike in one case but timelike in the
other.  However, this discontinuity lies in the metric; the resulting normal
vectors, taken together, form a continuous vector field.  Since the volume
element is just the (wedge) product of the (dual) frame elements, it, too,
must be continuous.  As mentioned above (and discussed in more detail
in~\cite{Junction}), it is enough for the pullbacks of the metric from $M^\pm$
to $\Sigma$ to agree in order for there to be a well-defined tangent space
on~$M$, a condition which is satisfied by this construction.

As discussed in~\cite{Failure}, Israel's results~\cite{Israel} relating the
stress-energy tensor to the intrinsic and extrinsic curvature of the boundary
layer $\Sigma$ must be modified in the presence of signature change.  For
example, the ``energy'' density on $\Sigma\subset M^\pm$ is now given by
\begin{equation}
\rho
  := G_{ab} n^a n^b
   = \frac12 \left( (K^c{}_c)^2 - K_{ab} K^{ab} - \epsilon R \right)
\end{equation}
\index{Darmois junction conditions}
where $K_{ab}$ is the extrinsic curvature of $\Sigma$, $R$ is the scalar
curvature of $\Sigma$, and $\epsilon=n_a n^a=\pm1$.  Imposing Darmois boundary
conditions, the curvatures themselves are continuous---but $\epsilon$ is not.
Thus, rather than the Israel condition $[\rho]=0$, we obtain
\begin{equation}
[\rho] = \left[G_{ab}n^an^b\right] = -R .
\end{equation}
Furthermore, we can independently recover the extrinsic curvature term from
\begin{equation}
\left[ G^a{}_b n^b l_a \right] = (K^c{}_c)^2 - K_{ab} K^{ab}
\end{equation}
where $l_a$ is the dual vector satisfying $l_an^a=1$ (on both sides), as using
$l_a$ instead of $n_a$ is equivalent to adding a factor of $\epsilon$ inside
the square brackets.  How to interpret ``energy'' inside a spacelike region
is, of course, an open question.

\section{Lorentzian Vector Calculus}
\label{TDcalc}

\index{vector calculus}%
A differential geometer regards vector calculus as ``really'' being about
differential forms.  For the last 20 years, in addition to my traditional
research in relativity, I have had the pleasure of attempting to implement an
approach to the teaching of vector calculus which can be described as
``differential forms without differential forms''~\cite{Bridge,BridgeBook}.
Key to this description is the use of the infinitesimal displacement vector
$d\rr$, which is really a vector-valued differential form.
\footnote{My recent textbook on general relativity~\cite{DFGGR} also uses this
language.}

When moving from Euclidean geometry to Riemannian geometry to Lorentzian
geometry, vector calculus as expressed in terms of differential forms is
virtually unchanged.  However, the conversion to traditional vector language
does depend on the signature.  It's easy to rewrite the dot product in
Lorentzian signature; relativists do this routinely when working with
Lorentzian metrics.  However, it's less obvious whether there is a cross
product, or what it looks like.  But nowhere is this dependence on signature
more apparent than in the statement of the Divergence Theorem.  What is an
``outward-pointing'' vector field, anyway?

In order to make the analogy to vector calculus more apparent, let's work in
$2+1$-dimensional Minkowski space, with orthonormal basis vectors $\Hat{x}$,
$\Hat{y}$, and $\Hat{t}$.  This basis satisfies
\begin{equation}
\Hat{x}\cdot\Hat{x} = 1 = \Hat{y}\cdot\Hat{y},
\qquad
\Hat{t}\cdot\Hat{t} = -1,
\end{equation}
with all cross terms vanishing.  So consider a vector field
\begin{equation}
\Vec{F} = F^x\,\Hat{x} + F^y\,\Hat{y} + F^t\,\Hat{t} .
\end{equation}
In Minkowski space, the divergence of $\Vec{F}$ is
\begin{equation}
\Vec\nabla\cdot\Vec{F}
  = \frac{\partial F^x}{\partial x} + \frac{\partial F^y}{\partial y}
	+ \frac{\partial F^t}{\partial t} ;
\end{equation}
there are no minus signs.  As in the component-based proof of the (ordinary)
Divergence Theorem, we can integrate the divergence over a rectangular box
$W$.  The first term yields
\begin{equation}
\int\!\int\!\int_W \frac{\partial F^x}{\partial x}\>\>dx\,dy\,dt
  = \int\!\int\Delta F^x\>\>dy\,dt
  = \int_{S_x} \Vec{F}\cdot\Hat{n}\>\>dA
\end{equation}
where $S_x$ consists of the two faces of the box with outward-pointing normal
vectors $\Hat{n}=\pm\Hat{x}$.  A similar expression holds for the
$y$-component, but the last equality fails for the $t$ component, since the
dot product has the wrong sign.  To fix this problem, we must instead choose
$\Hat{n}$ to be the \textit{inward}-pointing normal vector on $S_t$.

Why this asymmetry?  Stokes' Theorem~(\ref{Stokes}) is really about
differential forms, and the 1-form physically equivalent to $\Hat{n}$ (with
components $n_a$) is outward-pointing on all of $S$.  The ``asymmetry'' arises
due to the metric when converting from 1-forms to vectors.

Why haven't we noticed this asymmetry in relativity?  In practice, the
Divergence Theorem is not applied to closed regions $W$, but rather to
infinite ``sandwiches'', the region between two spacelike hypersurfaces.  The
integrals over the timelike sides of the box are replaced by falloff
conditions at spatial infinity, leaving only the ``$S_t$'' contribution in the
above argument.  The relative sign difference on spacelike and timelike
boundaries becomes an overall sign, which can be---and is---safely ignored.

The computations in this section are straightforward, but at first sight the
conclusion may be uncomfortable to some readers.  The Lorentzian Divergence
Theorem does \textit{not}, in general, involve the outward-pointing normal
vector (but rather the outward-pointing 1-form).  It is precisely this sort of
confrontation between expectation and reality that leads students to an
enhanced understanding of the underlying mathematics even in the traditional
setting.

The investigation of the Divergence Theorem for regions in Minkowski space
whose boundaries contain null pieces is left as an exercise for the reader.

\section{Summary}
\label{TDsum}

We have briefly summarized two quite different bodies of work, the analysis of
shells of matter in black-hole spacetimes, and of signature-changing
spacetimes, emphasizing the common thread provided by the Patchwork Divergence
Theorem, which also sheds new insight on topics in vector calculus.  One never
knows where the journey will lead.

Thank you, Paddy, for your contributions to my journey.

\section*{Acknowledgments}

\vspace{-0.1in}
It is a pleasure to thank the many collaborators who contributed to my work on
piecewise smooth manifolds, including especially Chris Clarke, George Ellis,
Charles Hellaby, Gerard 't Hooft, Corinne Manogue, and Robin Tucker.
\vspace{-0.1in}

%\section*{Appendix}
%\addcontentsline{toc}{section}{Appendix}

%\makeatletter
%\renewcommand\@biblabel[1]{#1.}
%\makeatother
%\bibliographystyle{plain}
%\bibliography{paddy}

%\printindex

\end{document}